\documentclass[final,1p]{elsarticle} 

\usepackage{lineno,hyperref}
\usepackage{amsmath}

\begin{document}

\begin{frontmatter}

\title{Improved muon decay simulation with \textsc{McMule} and \textsc{Geant4}}

\author[PaviaU,PaviaINFN]{A.~Gurgone\corref{AG}}
\cortext[AG]{Corresponding author.}
\ead{andrea.gurgone01@ateneopv.it}

\author[PisaU,PisaINFN,PSI]{A.~Papa}
\author[Washington]{P.~Schwendimann}
\author[PSI,UZH]{A.~Signer}
\author[IPPP]{Y.~Ulrich}

\author[PisaU,PisaINFN]{A.~M.~Baldini}
\author[PisaU,PisaINFN]{F.~Cei}
\author[PisaINFN]{M.~Chiappini}
\author[PisaU,PisaINFN]{M.~Francesconi}
\author[PisaINFN]{L.~Galli}
\author[PisaINFN]{M.~Grassi}
\author[PisaU,PisaINFN]{D.~Nicol\`o}
\author[PisaINFN]{G.~Signorelli}

\address[PaviaU]{Dipartimento di Fisica, Universit\`a di Pavia, Via Agostino Bassi 6, 27100 Pavia, Italy}
\address[PaviaINFN]{INFN, Sezione di Pavia, Via Agostino Bassi 6, 27100 Pavia, Italy}
\address[PisaU]{Dipartimento di Fisica, Universit\`a di Pisa, Largo Bruno Pontecorvo 3, 56127 Pisa, Italy}
\address[PisaINFN]{INFN, Sezione di Pisa, Largo Bruno Pontecorvo 3, 56127 Pisa, Italy}
\address[PSI]{Paul Scherrer Institut, Forschungsstrasse 111, 5232 Villigen, Switzerland}
\address[Washington]{Department of Physics, University of Washington, Box 351560, Seattle, Washington 98195, USA}
\address[UZH]{Physik-Institut, Universit\"at Z\"urich, Winterthurerstrasse 190, 8057 Z\"urich, Switzerland}
\address[IPPP]{Institute for Particle Physics Phenomenology, University of Durham, Durham DH1 3LE, United Kingdom}

\begin{abstract}
The physics programme of the MEG~II experiment can be extended with the search for new invisible particles produced in rare muon decays.
The hunt for such elusive signals requires accurate simulations to characterise the detector response and estimate the experimental sensitivity.
This work presents an improved simulation of muon decay in MEG~II, based on \textsc{McMule} and \textsc{Geant4}.
\end{abstract}

\begin{keyword}
Muon decay \sep Axion-like particles \sep MEG~II \sep Tracking detectors \sep Monte Carlo simulations
\end{keyword}

\end{frontmatter}


\section{Introduction}

The search for charged Lepton Flavour Violation~(cLFV) in muon decays is a key tool to probe the Standard Model.
The MEG~II experiment at the Paul Scherrer Institut~(PSI) searches for ${\mu^+\to e^+\gamma}$ with a sensitivity on the branching ratio of ${6\cdot10^{-14}}$ at 90\% confidence level~\cite{MEGII:2018kmf}.
The experiment is also competitive in searching for muon decays involving a light neutral scalar boson $X$, such as an axion-like particle~(ALP).
Since MEG~II is designed for a two-body signal, a feasible process is ${\mu^+\to e^+X}$, whose signature is a monochromatic positron close to the kinematic endpoint of the ${\mu^+\to e^+\nu_e\bar\nu_\mu}$ background~\cite{Gurgone:2021mqd}.
The hunt for such an elusive signal requires exhaustive Monte Carlo~(MC) simulations to characterise the detector response and estimate the experimental sensitivity.
Since the radiative corrections at the endpoint are enhanced by the emission of soft photons, the simulation must include extremely accurate theoretical predictions for the event generation of both decays.


\section{Theoretical input}

\begin{figure}[ht!]
    \centering
    \includegraphics[width=.62\linewidth]{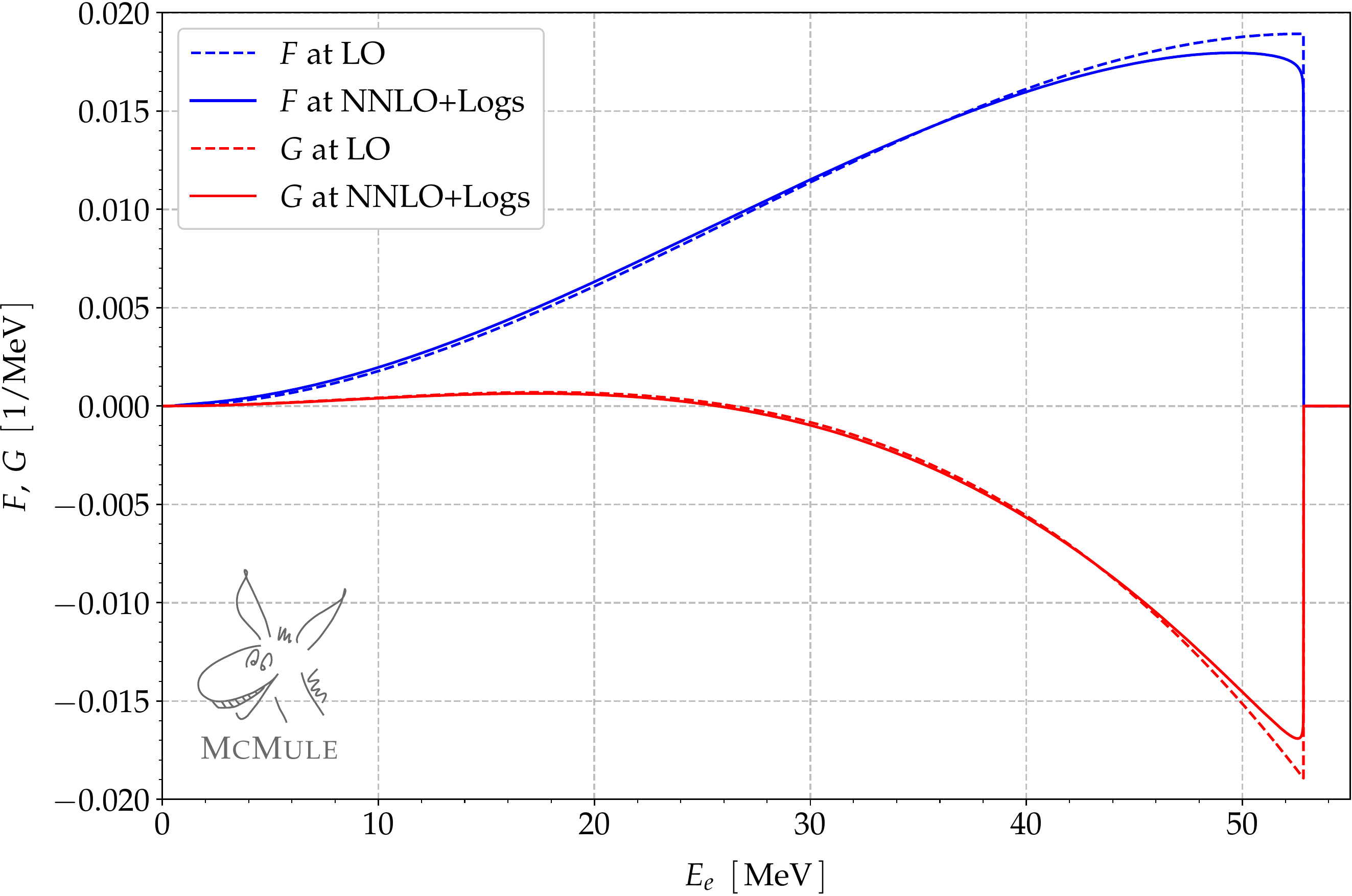}
    \caption{The functions $F$ and $G$ for ${\mu^+\to e^+\nu_e \bar\nu_\mu}$ computed with \textsc{McMule}.}
    \label{fig:theory}
\end{figure}

The needed accuracy is achieved with \textsc{McMule}, a numerical framework for the computation of fully differential QED corrections for low-energy processes with leptons~\cite{Banerjee:2020rww}.
For both decays, the relevant observables are the positron energy $E_e$ and the angle $\theta_e$ between the positron momentum $\vec{p}_e$ and the muon polarisation $\vec{n}_\mu$. 
The differential decay width can be written as
\begin{linenomath*}
\begin{equation}
\label{eq:fg}
\frac{\textrm{d}^2\Gamma}{\textrm{d}E_e\,\textrm{d}\hspace{-1pt}\cos\theta_e} = \frac{G_F^2 \, m_\mu^5}{192 \, \pi^3}\, \Big[ F(E_e) + n_\mu \cos\theta_e \, G(E_e) \Big]
\end{equation}
\end{linenomath*}
where $G_F$ is the Fermi constant and $m_\mu$ the muon mass. 
The two functions $F$ and $G$ contain all the information required to generate inclusive positron events.
In \textsc{McMule} the signal~${\mu^+\to e^+X}$ is implemented assuming a generic mass and coupling for the ALP and including the QED corrections at next-to-leading order~(NLO).
The background~${\mu^+\to e^+\nu_e \bar\nu_\mu}$ (Fig.~\ref{fig:theory}) includes the leading weak and hadronic corrections, exact QED corrections at next-to-next-leading order~(NNLO) and approximated logarithmically enhanced terms at higher orders~\cite{Banerjee:2022nbr}.

\section{Event generation}

\begin{figure}[ht!]
    \centering
    \includegraphics[width=.4\linewidth]{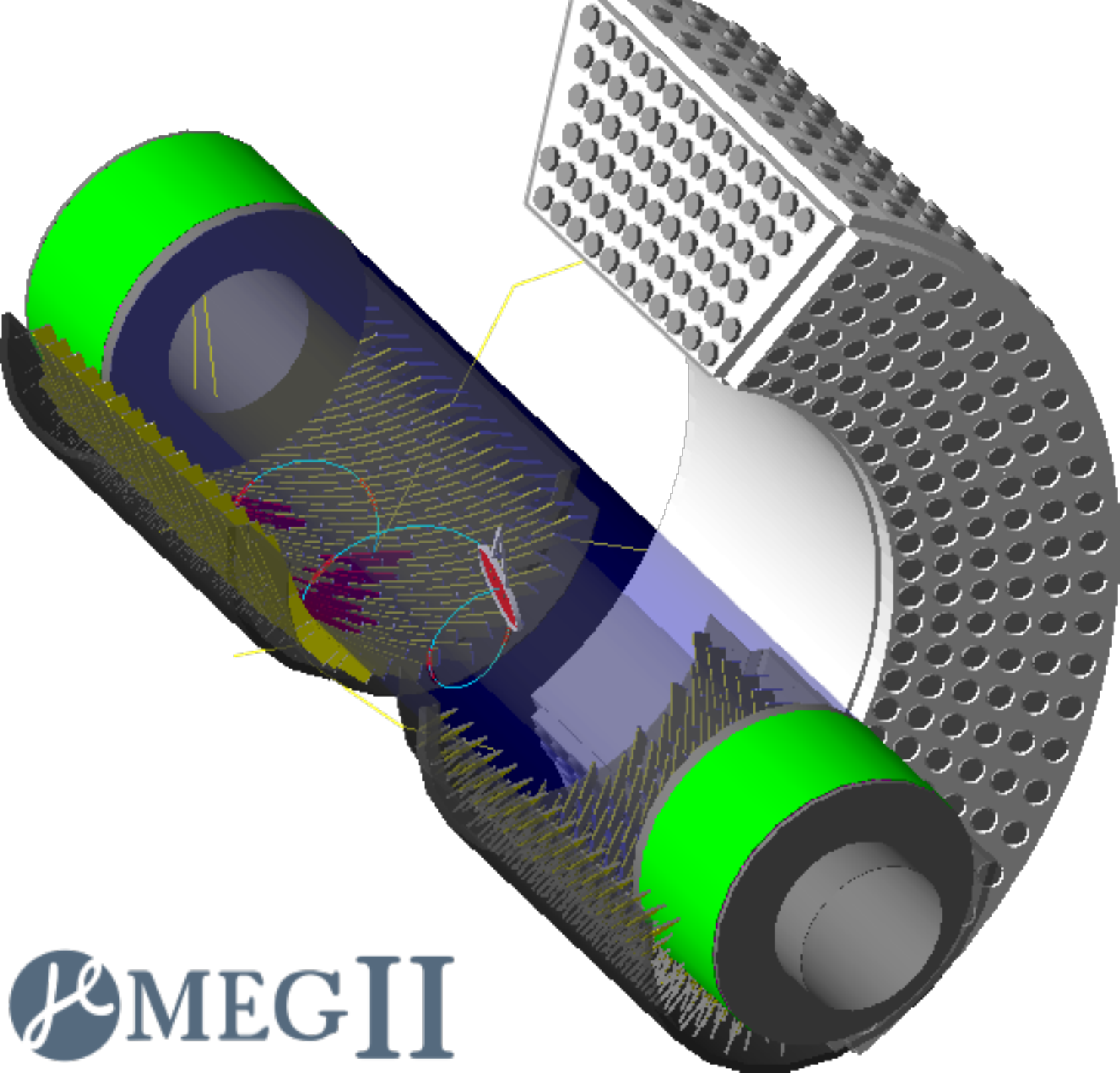}
    \caption{Simulation of a ${\mu^+ \to e^+ X}$ event in the MEG~II software.}
    \label{fig:event}
\end{figure}

The \textsc{McMule} predictions are used to implement a new positron event generator in the MEG~II software (Fig.~\ref{fig:event}).
The incoming muons are assumed to decay at rest inside the MEG~II target with a residual 85\% polarisation~\cite{MEGII:2018kmf}.
The decay vertex distribution is then obtained by intersecting the target geometry with the measured muon beam spot. 
The positron energy~$E_e$ and the polar angle~$\theta_e$ are generated accordingly to~\eqref{eq:fg}, while the azimuthal angle~$\phi_e$ is generated uniformly.
This set of observables completely specifies the event~kinematics.\\

\section{Positron reconstruction}

\begin{figure}[ht!]
    \centering
    \includegraphics[width=.635\linewidth]{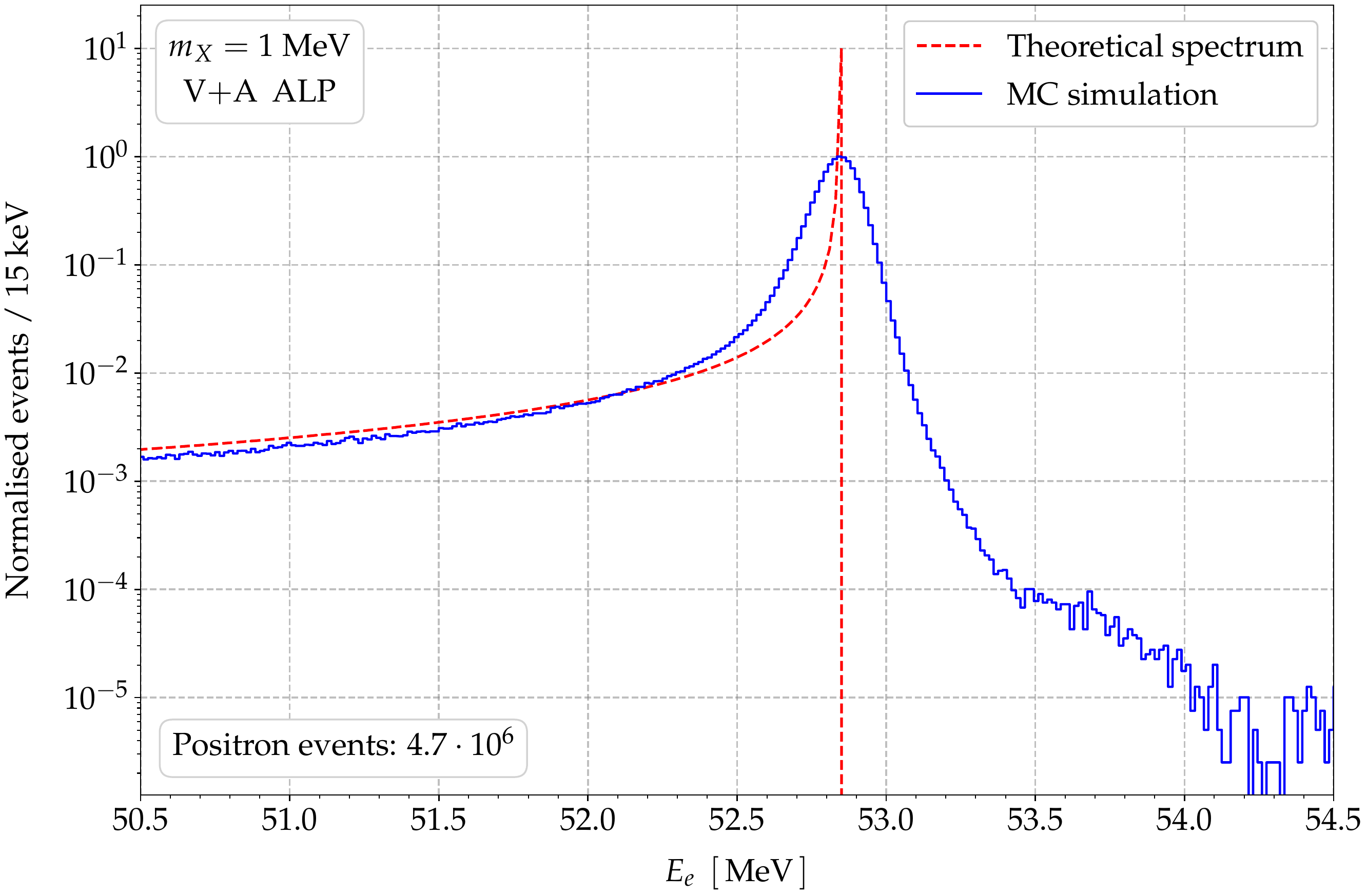}
    \caption{Simulation of positron energy reconstruction for ${\mu^+ \to e^+ X}$.}
    \label{fig:hsig}
\end{figure}

\begin{figure}[ht!]
    \centering
    \hspace{1mm}
    \includegraphics[width=.62\linewidth]{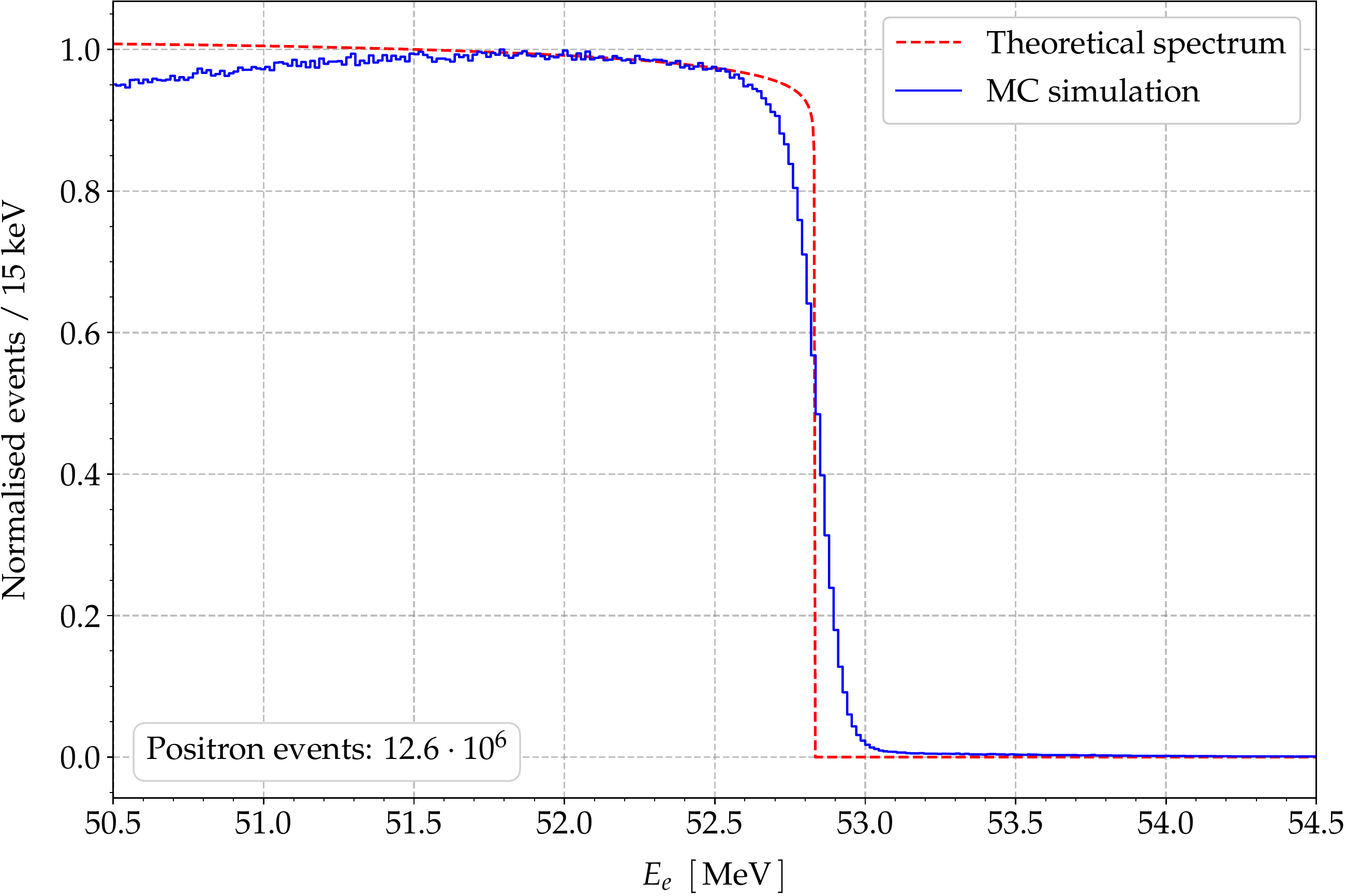}
    \caption{Simulation of positron energy reconstruction for ${\mu^+\to e^+\nu_e \bar\nu_\mu}$.}
    \label{fig:hbgr}
\end{figure}

In MEG~II the outgoing positrons are tracked with a spectrometer consisting of three main elements:
\begin{enumerate}
    \item The COnstant Bending RAdius~(COBRA) magnet, a solenoid with a field gradient along the beam direction.
    \item The Cylindrical Drift CHamber~(CDCH) for the measurement of the decay vertex position (${\delta x_\mu \approx 1}$~mm) and the positron momentum (${\delta p_e \approx 100}$~keV).
    \item The pixelated Timing Counter~(pTC) for the measurement of the positron emission time (${\delta t_e \approx 35}$~ps).
\end{enumerate}
The positron reconstruction is simulated with a dedicated software.
After the event generation, the detector hits are simulated with \textsc{Geant4}~\cite{GEANT4:2002zbu} and converted into electronic signals.
In order to extract the physical observables, the resulting waveforms are converted into raw data, such as signal charge and time.
The positron tracks are identified with a pattern recognition algorithm and fitted with a Kalman filter, applying the same procedure used for real data.
Their quality is ensured by requiring less than three turns in the spectrometer and at least 25 hits in the CDCH and one hit in the pTC.
The results for both ${\mu^+ \to e^+ X}$ (Fig.~\ref{fig:hsig}) and ${\mu^+\to e^+\nu_e \bar\nu_\mu}$ (Fig.~\ref{fig:hbgr}) are~reported.\\

\section{Conclusion and outlook}

\begin{figure}[ht!]
    \centering
    \includegraphics[width=.63\linewidth]{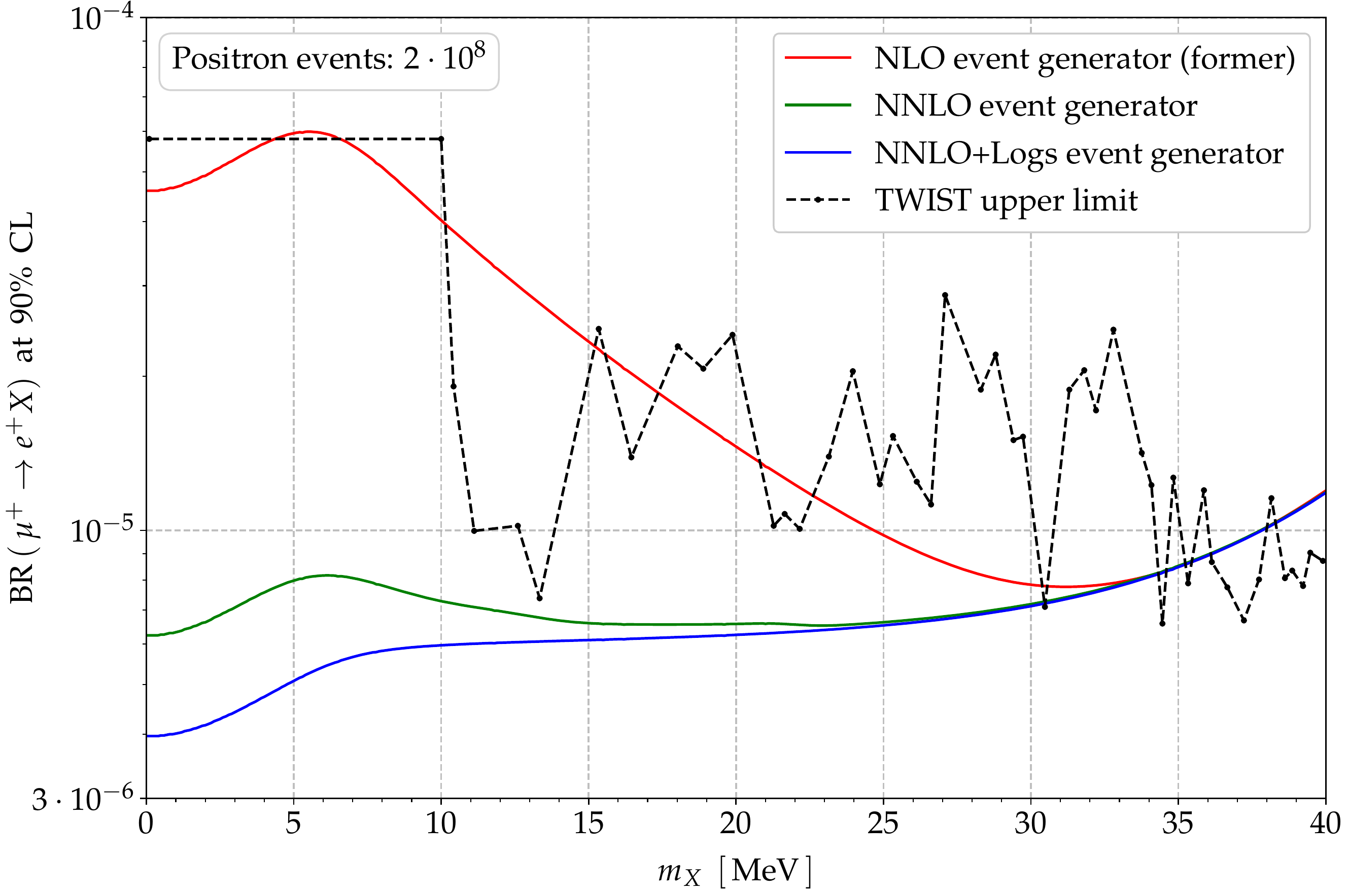}\hspace{7pt}
    \caption{Expected sensitivity on ${\mu^+ \to e^+ X}$ for different implementations of the positron event generator, showing the effect of the reduced theoretical error.}
    \label{fig:sens}
\end{figure}

The search for flavour-violating ALPs in muon decays such as ${\mu^+ \to e^+ X}$, ${\mu^+ \to e^+ X \gamma}$ or ${\mu^+ \to e^+ (X \to \gamma \gamma)}$ is an excellent opportunity to extend the MEG~II physics programme beyond ${\mu^+ \to e^+ \gamma}$.
Although ${\mu^+ \to e^+ X}$ is particularly elusive, the MEG~II spectrometer is specifically designed for tracking positrons at the endpoint,\linebreak\\
as occurs for ${m_X\simeq0}$.
A preliminary study, based on a cut-and-count approach~\cite{Gurgone:2021mqd,Banerjee:2022nbr}, shows a competitive sensitivity (Fig.~\ref{fig:sens}), close to the upper limit set by TWIST~\cite{TWIST:2014ymv}.
However, since an offset on the positron energy scale results in a false signal at the endpoint, a rigorous control of the systematic effects is required.
To this end, new calibration tools for the positron spectrometer are under development.\\

\bibliography{bibliography}

\end{document}